\documentclass[aps,prd,groupedaddress,showpacs,notitlepage]{revtex4-2}
\usepackage{amsmath,amstext,amsbsy,amssymb}
\usepackage{bm}
\usepackage{color}

\usepackage[linktocpage]{hyperref}

\newcommand{\cross}{\times}

	\newcommand{\ssA}{{\scriptscriptstyle{ A}}}
	\newcommand{\ssV}{{\scriptscriptstyle{ V}}}

	\newcommand{\ssR}{{\scriptscriptstyle{ R}}}

	\newcommand{\ssE}{{\scriptscriptstyle{ E}}}
	
	\newcommand{\ssL}{{\scriptscriptstyle{ L}}}
	
	\newcommand{\ssM}{{\scriptscriptstyle{ M}}}
	\newcommand{\ssC}{{\scriptscriptstyle{ C}}}

	\long\def\symbolfootnote[#1]#2{\begingroup%
		\def\thefootnote{\fnsymbol{footnote}}\footnote[#1]{#2}\endgroup}

\begin{document}
		
		\title{Nonrelativistic Transport Theory from  Vorticity Dependent Quantum Kinetic Equation }

		\author{\"{O}mer F. Dayi}
		\author{S. Efe Gürleyen}
		\affiliation{%
			Physics Engineering Department, Faculty of Science and
			Letters, Istanbul Technical University,
			TR-34469, Maslak--Istanbul, Turkey}

		\begin{abstract}
		
			We study the three-dimensional  transport theory of massive spin-1/2 fermions resulting from the vorticity dependent quantum kinetic equation.
			This quantum kinetic equation  has been introduced to take account of noninertial properties of  rotating coordinate frames. We show that it is the appropriate relativistic kinetic equation which provides  the vorticity dependent semiclassical transport equations of the  three-dimensional Wigner function components. 
			We establish  the semiclassical kinetic equations of a linearly independent set of  components.
			By means of them,  kinetic equations of the chiral scalar distribution functions are derived.  They furnish the 3D kinetic theory which permits us to study  the vector and axial vector current densities by focusing on the mass corrections to the chiral vortical and separation effects.

		\end{abstract}
		
		\maketitle
		\section{Introduction}
		
		Transport of Dirac fermions in the presence of  external  electromagnetic fields can be studied by means of the covariant  Wigner function which obeys  the quantum kinetic equation (QKE) \cite{egv,vge}.   The
		Wigner function can be decomposed into  some  covariant fields whose equations of motion  follow from the QKE. 
		Founded on these field equations, one derives  relativistic transport theories of Dirac particles.  A brief overview of the covariant Wigner function approach was given  in \cite{RecRev} and recently  it has  been reviewed  in details in  \cite{ hidakaRev}. Relativistic  formalism has the advantage  of being manifestly Lorenz invariant. Nevertheless, nonrelativistic  transport equations are necessary for being able to start with  initial distribution functions  and construct solutions of transport equations  \cite{bbgr,zh2,oh}. There exist some different methods of formulating nonrelativistic  kinetic theories of Dirac particles. One of these methods is to  construct the  four-dimensional (4D) transport equations of a set of  covariant fields and then integrate them over the zeroth-component of four-momentum, so that  the three-dimensional (3D)   transport equations which are correlative to the 4D  ones are extracted.  Another method  is to  integrate all of the quantum kinetic equations of the covariant fields over the zeroth-component of four-momentum  at the beginning and then   derive the  nonrelativistic  transport theory from  these 3D quantum kinetic equations \cite{zh1,oh}.  This is  also called the equal-time formalism.
		There also exists a strictly 3D approach of acquiring a transport equation of Dirac particles which does not refer to the Wigner function \cite{dky}.  
		
		Quarks of the quark-gluon plasma formed in heavy-ion collisions are treated as massless \cite{Gyulassy:2004zy, Shuryak:2004cy}. Thus, chiral kinetic theory (CKT) is useful to inspect their dynamical features \cite{kmw,fkw,kz,mz,jkr,ss,lw,bpr,glpww}.
The QKE of the  relativistic Wigner function generates the anomalous magnetic effects as well as the vorticity effects correctly \cite{glpww}.   It is worth noting that  the vorticity of fluid  matches the angular velocity of the fluid in the comoving  frame. The QKE  possesses an explicit dependence on the electromagnetic fields but  not on the  vorticity of fluid. 
		When the Wigner function is expressed in the Clifford algebra basis, the  QKE  gives a set of equations for the chiral vector fields. In solving some of these equations one introduces the  frame four-vector \cite{hpy1,hpy2,hsjlz}. It  can be identified with  the comoving frame velocity which also appears  in equilibrium distribution functions. Derivatives of  the comoving frame four-velocity generate  terms depending on the vorticity. These are the sources of  vorticity  dependence in the relativistic QKE formulations of the massless fermions.     One cannot generate noninertial  forces like the Coriolis force within this formalism.   However, in \cite{SteYin}, vortical effects were derived by using the similarity between  the Lorentz and the Coriolis forces. This formulation has been shown to result in a rotating coordinate frame from the first principles \cite{husad}. To build in this similarity, a modification of QKE by means of enthalpy current was introduced in \cite{dk}. The discrepancy in treating magnetic and vortical effects reflects itself  drastically,  especially in 3D CKT when both electromagnetic fields and vorticity are taken into account.  The vorticity dependent quantum kinetic equation (VQKE) was shown to yield a 3D CKT which does not depend on the spatial coordinates  \cite{dk}. It is consistent with the chiral anomaly and  generates the chiral  magnetic and vortical effects and the Coriolis force.   The underlying Lagrangian  formalism  which yields  VQKE was presented in \cite{bizson}. 
		
		Constituent quarks  of the quark-gluon plasma created  in heavy-ion collisions are approximately massless. Thus, to get a better understanding of their dynamical properties, one needs to uncover the mass corrections to  chiral theories.  Covariant kinetic theories of  massive spin-1/2 particles  have been studied in terms of  QKE within two different approaches in \cite{wsswr,hhy}.   In principle,  nonrelativistic transport equations  can be provided by  integrating the 4D kinetic equations. But, for massive fermions, extracting  the 3D kinetic  equations which are correlative to the 4D kinetic equations  can only be done under some simplifying approximations  as they have been shown within the  VQKE approach in \cite{dk-m,bizson}.   
		We  have already mentioned that there also exists another nonrelativistic approach which is  the so-called equal-time formalism. By integrating the  equations of the relativistic Wigner function components  over the zeroth-component of momentum, one sets up the  equations of the components of the 3D Wigner function. Then, one employs these 3D equations to derive  nonrelativistic kinetic equations of Dirac particles in the presence of the external electromagnetic fields. This has been studied in \cite{masscorr}  by developing the original formulation of \cite{zh1}.  In contrast to the 4D Wigner function approaches, this 3D formalism does not generate vortical effects. Because, without solving some of the equations of the 4D Wigner function components one cannot generate vorticity dependent terms.    Therefore, to obtain a similar 3D  approach  by taking account of vorticity of the fluid, the QKE of the Wigner function should possess an explicit dependence on the fluid  vorticity.  The VQKE is the unique covariant formalism which has this property \footnote{An attempt  to study nonrelativistic kinetic theory of massive fermions  in the presence of rotational field within the 3D Wigner function method is presented in \cite{cpl}.  However, they  start with a wave equation which is not covariant. For us it is obscure how one can justify use of the covariant Wigner function when it is constructed by a wave function which obeys a non-covariant equation of motion. }. In this work, we  study  the  3D formulation of the VQKE by extending  the method of \cite{zh1,masscorr}, by taking care of  the vortical effects only. 
		
		We briefly review the VQKE in the absence of electromagnetic fields and present  the  equations of the components of the covariant Wigner function in the next section. Their integration over the zeroth component of  momentum in a  frame adequate to study nonrelativistic dynamics lead to the 3D constraint and transport  equations as reported in Sec. \ref{3Dcte}. These equations which the components of  the 3D  Wigner function  obey are studied
		  and their  semiclassical solutions are acquired in terms of a set of independent functions.  In Sec. \ref{skefg}, kinetic equations of this set of fields are established   up to the first order in   the Planck constant, $\hbar .$ Mass corrections to the chiral vortical and separation effects  are studied in Sec. \ref{masKT}. Conclusions  and discussions of  possible future directions are given in Sec. \ref{disc}.
		
		\section{Vorticity Dependent Quantum Kinetic Equation}
		\label{secvqke}
		
		The quantum kinetic equation for a fluid in the comoving  frame with the four-velocity $u_\mu;$ $u_\mu u^\mu =1,$  whose linear acceleration vanishes, $ u_\nu \partial^\nu u_\mu =0,$ is 
		\begin{equation}
			\left[\gamma_\mu \left(\pi^\mu + \frac{i\hbar}{2} 	{\mathcal D}^\mu \right)-m \right] W(x,p) = 0.
			\label{qke}
		\end{equation}
		Here,
		\begin{eqnarray}
			{\mathcal D}^{\mu} &\equiv & \partial^{\mu}-j_{0}(\Delta) w^{\mu\nu}   \partial_{p \nu} , \label{Df}\\
			\pi^{\mu} &\equiv &p^{\mu}-\frac{\hbar}{2} j_{1}(\Delta)w^{\mu\nu} \partial_{p \nu} ,\label{Pf}
		\end{eqnarray}
		where $\partial^\mu \equiv \partial / \partial x_\mu,$  $\partial_p^\mu \equiv \partial / \partial p_\mu,$ and $j_{0},j_{1}$
		are spherical Bessel functions in $\Delta \equiv \frac{\hbar}{2} \partial_{p} \cdot \partial_{x}.$ The 4D space-time derivative, \(\partial_{\mu},\) contained in \(\Delta\) acts on \(w^{\mu\nu}  ,\) but not on the Wigner function, $W(x,p).$  On the contrary,  $ \partial_{p \nu}$ acts on the Wigner function, but not on  \( w^{\mu\nu}  .\)  The action which generates \eqref{qke} has been presented in \cite{bizson}. There, it was shown that when the equations of motion of the fields presenting fluid are satisfied, $w^{\mu\nu} $ is given with an arbitrary constant $\kappa$ as
		\begin{eqnarray}
			w_{\mu\nu} =(\partial_\mu h)u_\nu - (\partial_\nu h)u_\mu  + \kappa  h\Omega_{\mu \nu},
			\label{w_munu}
		\end{eqnarray}
		where $h=u\cdot p$ and   $\Omega_{\mu\nu} =\frac{1}{2}( \partial_\mu u_\nu -\partial_\nu u_\mu ).$ The fluid four-vorticity is defined as $\omega_\mu=(1/2)\epsilon_{\mu \nu \alpha \beta}u^\nu \Omega^{\alpha \beta} .$

		The  Wigner function can be written through the 16 generators of the Clifford algebra as
		\begin{equation}
			W=\frac{1}{4}\left(\mathcal{F}+i \gamma^{5} \mathcal{P}+\gamma^{\mu} \mathcal{V}_{\mu}+\gamma^{5} \gamma^{\mu} \mathcal{A}_{\mu}+\frac{1}{2} \sigma^{\mu \nu} \mathcal{S}_{\mu \nu}\right),
			\label{wigner}
		\end{equation}
		where  $\mathcal{C}_a \equiv\left\{ \mathcal{F},\mathcal{P},\mathcal{V}_{\mu},\mathcal{A}_{\mu},\mathcal{S}_{\mu \nu}\right\} ,$  respectively, are the scalar, pseudoscalar, vector, axial-vector, and antisymmetric tensor components of the 4D Wigner function.   
		These covariant fields can be expanded in powers of  the Planck constant:
		\begin{equation}
			\mathcal{C}_a =\sum_n\hbar^n\mathcal{C}^{(n)}_a.
		\end{equation} 
		We  deal with the semiclassical approximation where only the zeroth- and first-order fields in $\hbar$ are considered. Thus, to derive the equations which they satisfy, instead of  (\ref{Df}), (\ref{Pf}), we only need to deal with 
		\begin{eqnarray}
			D^{\mu} &\equiv & \partial_{x}^{\mu}-w^{\mu\nu}  \partial_{p \nu}  , \label{dmu}
		\end{eqnarray}
	and $p^{\mu} . $
		By  plugging   the decomposed Wigner function, \eqref{wigner}, into the VQKE, (\ref{qke}),  one  derives the  equations satisfied by the  fields $\mathcal{C}_a , $ whose  real parts are
		\begin{eqnarray}
			p\cdot \mathcal{V}-m \mathcal{F} =0,  \label{real1} \\
			{p_{\mu} \mathcal{F}-\frac{\hbar}{2} D^{\nu} \mathcal{S}_{\nu \mu}-m \mathcal{V}_{\mu}=0},  \label{real2} \\
			{-\frac{\hbar}{2} D_{\mu} \mathcal{P}+\frac{1}{2} \epsilon_{\mu \nu \alpha \beta} p^{\nu} S^{\alpha \beta}+m \mathcal{A}_{\mu}=0}, \label{real3} \\
			{\frac{\hbar}{2} D_{[\mu} \mathcal{V}_{\nu]}-\epsilon_{\mu \nu \alpha \beta} p^{\alpha} \mathcal{A}^{\beta}-m \mathcal{S}_{\mu \nu}=0}, \label{real4} \\
			\frac{\hbar}{2} D \cdot \mathcal{A}+m \mathcal{P} =0, 
			\label{real5}
		\end{eqnarray}
		and the imaginary parts are
		\begin{eqnarray}
			{\hbar D \cdot \mathcal{V}=0}, \label{imag1} \\ 
			{p \cdot \mathcal{A}=0}, \label{imag2} \\ 
			{\frac{\hbar}{2} D_{\mu} \mathcal{F}+p^{\nu} \mathcal{S}_{\nu \mu}=0},  \label{imag3} \\ 
			{p_{\mu} \mathcal{P}+\frac{\hbar}{4} \epsilon_{\mu \nu \alpha \beta} D^{\nu} \mathcal{S}^{\alpha \beta}=0},  \label{imag4} \\ 
			{p_{[\mu} \mathcal{V}_{\nu]}+\frac{\hbar}{2} \epsilon_{\mu \nu \alpha \beta} D^{\alpha} \mathcal{A}^{\beta}=0}.
			\label{imag5}
		\end{eqnarray}
		It can be observed that not all of  the fields ${\cal C}_a$ are relevant to formulate a 4D kinetic theory.  Depending on the choice of the independent set of fields, one establishes different relativistic kinetic theories \cite{hhy,wsswr}.  In the subsequent sections  we will refer to the  4D formulation given in \cite{dk-m}, which was acquired following the approach of \cite{wsswr}.

		\section{3D semiclassical transport and constraint equations }
		\label{3Dcte}
		
		The equal-time transport theory of Dirac particles in the presence of external electromagnetic fields which has been  proposed in \cite{bbgr} was incomplete. In  \cite{zh1}, it was shown that to have a complete  nonrelativistic transport theory of spinor electrodynamics, one should start with the covariant QKE of \cite{egv,vge}.
		We mainly adopt the approach of  \cite{zh1}. However, there is a subtle difference: Electromagnetic field strength is independent of $p_0$ in contrast to $w_{\mu \nu} $ which explicitly depends on it. We will show how to surmount this difficulty.
		
		The nonrelativistic (3D) Wigner function is defined as the integral of the 4D Wigner function over the zeroth-component of momentum:
		\begin{equation}\label{3d Wigner exp}
			W_3(x,\bm p)=\int dp_0 W(x,p)\gamma_0.
		\end{equation}
		Let us define the 3D components in the Clifford algebra basis as 
		\begin{equation}
			W_3(x,\bm p)=\frac{1}{4}[f_0+\gamma_5 f_1-i\gamma_0 \gamma_5 f_2 +\gamma_0 f_3 +\gamma_5\gamma_0 \bm\gamma\cdot\mathbf{g}_0+\gamma_0\bm\gamma\cdot\mathbf{g}_1-i\bm\gamma\cdot\mathbf{g}_2-\gamma_5\bm\gamma\cdot\mathbf{g}_3].
		\end{equation}
		The 4D and 3D  components are related  as 
		\begin{equation}
			\begin{array}{lclclcl}
			f_0(x,\bm p) &= & \int dp_0 {\cal V}_0(x,p), &  &f_1(x,\bm p)  &= &  \int dp_0 {\cal A}_0(x,p), \\
			&&&&&\\
			f_2(x,\bm p)   &= &  \int dp_0 {\cal P}(x,p), &  &	f_3(x,\bm p)   &= &  \int dp_0 {\cal F}(x,p), \\
				&&&&&\\
			\mathbf{g}_0(x,\bm p) &= &  \int dp_0  \bm{{\cal A}}(x,p), &  & \mathbf{g}_1(x,\bm p) &= & \int dp_0 \bm{{\cal V}}(x,p),\\
				&&&&&\\
			\mathrm{g}_2^i(x,\bm p) &= & -\int dp_0 {\cal S}^{0i}(x,p), &  & \mathrm{g}_3^i(x,\bm p)  &= &\frac{1}{2}\epsilon^{ijk}\int dp_0 {\cal S}_{jk}(x,p). 
						\end{array}
		\nonumber
	\end{equation}
		
		To express $w_{\mu \nu}$  in terms of  the 3D vorticity, $\bm \omega,$ which is uniform,  we choose the  frame 
		\begin{equation}
			\begin{aligned}
				u^\mu=(1,\mathbf{0}), & &
				\omega^\mu=(0,\bm{\omega}).
			\end{aligned}
		\end{equation}
	Thus, \eqref{w_munu} yields
		$$
		w^{0i} = -\epsilon^{ijk}p_j \omega_k, \ 
		w^{ij} = \kappa p_0 \epsilon^{ijk}\omega_k.
		$$
		In contrast to the electromagnetic field strength, $w_{ij}$ is $p_0$ dependent.  
		In this frame, the components of $D_\mu=(D_t,\bm D),$ \eqref{dmu}, are
		\begin{eqnarray}
			D_t &=& \partial_t + (\bm p\times \bm{\omega})\cdot{\bm{\nabla}}_p, \label{Dt}  \\ 
			{\bm D} &=& \bm{\nabla}+\kappa p_0 \bm{\omega}\times\bm{\nabla}_p . \label{Dpo}
		\end{eqnarray}
		While  obtaining the 3D  formalism by integrating the 4D transport equations  over $p_0,$ the dependence of  $\bm D$  on  $p_0$   should be handled carefully.  
		
		To establish the 3D formalism, we integrate the relativistic  equations  \eqref{real1}-\eqref{imag5} over $p_0.$   They will be separated into two groups  \cite{zh1}  by inspecting their  dependence on the  time derivative $\partial_t.$  The equations containing $\partial_t$  yield  the transport equations:
		\begin{subequations}
			\begin{align}
				&	\hbar \left(D_t f_0 +\int dp_0 {\bm D}\cdot\mathbf{V}\right)  = 0,\label{8a}\\
				&	\hbar \left(D_t f_1 + \int dp_0 {\bm D}\cdot\mathbf{A}\right) +2mf_2  = 0,\label{8b}\\
				&	\hbar D_t f_2+2\bm p \cdot \mathbf{g}_3-2mf_1=0,\label{8c}\\
				&	\hbar D_t f_3-2\bm p \cdot \mathbf{g}_2 = 0,\label{8d}\\
				&	\hbar\left(D_t \mathbf{g}_0 + \int dp_0 {\bm D}A_0\right)-2 \bm {p}\times\mathbf{g}_1=0,\label{8e}\\
				&	\hbar\left(D_t \mathbf{g}_1 + \int dp_0 {\bm D}V_0\right)-2\bm{p}\times\mathbf{g}_0 + 2m\mathbf{g}_2 =0,\label{8f}\\
				&	\hbar \left( D_t {\mathrm{g}}^i_2 - \int dp_0 D_j S^{ji} \right) + 2p^i f_3-2m{\mathrm{g}}^i_1 =0,\label{8g}\\
				&	\hbar \left( D_t \mathrm{g}_3^i + \int dp_0 \varepsilon^{ijk} D_j S_{k0} \right) - 2p^i f_2 = 0.\label{8h}
			\end{align}
		\end{subequations}
		The others are  the  constraint equations:
		\begin{subequations}
			\begin{align}
				&	\int dp_0 p_0 V_0 -\bm{p}\cdot\mathbf{g}_1 - mf_3=0,\label{9a}\\
				&	\int dp_0 p_0 A_0 -\ \bm{p}\cdot\mathbf{g}_0 = 0,\label{9b}\\
				&	\int dp_0 p_0 P - \frac{1}{4}\hbar\int dp_0 \varepsilon_{ijk}D^i S^{jk} = 0,\label{9c}\\
				&	\int dp_0 p_0 F + \frac{1}{2}\hbar \int dp_0 D^i S_{0i}-mf_0=0,\label{9d}\\
				&	\int dp_0 p_0 \mathbf{A}-\bm{p}f_1-\frac{\hbar}{2}\int dp_0 {\bm D}\times\mathbf{V}-m\mathbf{g}_3 = 0,\label{9e}\\
				&	\int dp_0 p_0 \mathbf{V}-\bm{p}f_0-\frac{\hbar}{2}\int dp_0 {\bm D}\times\mathbf{A}=0,\label{9f}\\
				&	\int dp_0 p^0 S_{0i} -\bm p \times \mathbf{g}_3 + \frac{\hbar}{2}\int dp_0 D_i F=0,\label{9g}\\
				&	\frac{1}{2}\epsilon_{ijk}\int dp_0p^0 S^{jk}-(\bm p \times\mathbf{g}_2)_i-\frac{\hbar}{2}\int dp_0 D_i P+m\mathrm{g}_{0i}=0.\label{9h}
			\end{align}
		\end{subequations}
		These equations show that not all of the 3D fields are independent. In fact, we can express  them in terms of $f_0$ and $\mathbf{g}_0$ as it will be discussed  below.
		
		
		In the classical limit,  \eqref{qke} simplifies and yields the classical on-shell condition
		\begin{equation}
			(p^2-m^2) W(x,p) = 0, \label{ms0}
		\end{equation}
		whose solutions are $p_0=\pm E_p,$ where $E_p=\sqrt{\bm p^2+m^2}.$ Therefore, in the classical limit, the fields can be written as the sum of positive and negative energy solutions:
		\begin{equation}
			\mathcal{C} _a^{(0)}(x,p)=	\mathcal{C} _a^{(0)+}(x,p)\delta(p_0-E_p)+	\mathcal{C} _a^{(0)-}(x,p)\delta(p_0+E_p).
		\end{equation}
		Thus, at the  leading order in $\hbar,$ the $p_0$ integrals in  \eqref{8a}-\eqref{9h} can easily be performed and all of the 3D fields can be expressed in terms of $f_0$ and $\mathbf{g}_0$ in the classical limit   as \cite{zh1}
		\begin{eqnarray}
			f_1^{(0)\pm}&=&\pm \frac{\bm p}{E_p}\cdot\mathbf{g}_0^ {(0)\pm},\label{12a}\\
			f_2^{(0)\pm}&= &0,\label{12b}\\
			f_3^{(0)\pm}&= &\pm \frac{m}{E_p}f_0^{(0)\pm},\label{12c}\\
			\mathbf{g}_{1}^{(0)\pm}&= &\pm\frac{\bm p}{E_p}f_0^{(0)\pm},\label{12d}\\
			\mathbf{g}_2^{(0)\pm}&=& \frac{\bm p}{m}\cross\mathbf{g}_0^{(0)\pm},\label{12e}\\
			\mathbf{g}_3^{(0)\pm}&= &\pm\frac{E_p^2 \mathbf{g}_0^{(0)\pm}-(\bm p\cdot\mathbf{g}_0^{(0)\pm})\bm p}{mE_p}\label{12f}.
		\end{eqnarray}
		
		To solve the transport and constraint equations to determine the 3D fields in terms of $f_0$ and $\mathbf{g}_0$ at the $\hbar$ order, one can  attempt to add  a $\hbar$-order term to the on-shell condition  \eqref{ms0}. However, by inspecting the relativistic semiclassical solutions of  \eqref{real1}-\eqref{imag5} given in \cite{dk-m}, it can be observed that  each 
		field  ${\cal C}_a $ satisfies a different   mass shell condition at the  $\hbar$ order. In \cite{masscorr},  it was suggested to  write
		\begin{equation}\label{Continuous expansion}
			\mathcal{C} ^\pm_a(x,p) = \mathcal{C} _a^{(0)\pm }(x,p)\delta(p_0\mp E_p)+\hbar {\cal A}_a^\pm(p).
		\end{equation}
		and define the 3D shell shifts  as
		\begin{equation}\label{Continuous Shift}
			\Delta E_{a}^\pm (x,\bm p)= \int dp_0  p_0 {\cal A}_a^\pm(p).
		\end{equation}
		The operator $\bm D$ depends on $p_0,$ so that the  related energy averages are expressed as
		\begin{eqnarray}
			\int dp_0 {\bm D}{\cal C}_a(x.p) &=& \int dp_0 \left(\bm{\nabla}+\kappa p_0 \bm{\omega}\cross\bm{\nabla}_p \right)
			\left({\cal C}_a^\pm(x,p)\delta(p_0\mp E_p)+\hbar A_a^\pm(p) \right) \nonumber \\
			& =&   {\bm D}^{(0)}_\pm {\cal C}_a^\pm(x,\bm p )+	\hbar \kappa (\bm{\omega}\cross\bm{\nabla}_p) \Delta {E}_{a}^\pm , \label{DD}
		\end{eqnarray}
		where 
		\begin{eqnarray}
			{\bm D}^{(0)}_\pm&=&\bm{\nabla} \pm \kappa E_p   \bm{\omega}\cross\bm{\nabla}_p \pm\frac{\kappa}{E_p}\bm{\omega}\cross\bm p \nonumber\\
			&\equiv& \bm{\partial}_\pm^{(0)}\pm\frac{\kappa}{E_p}\bm{\omega}\cross\bm p. \label{D0}
		\end{eqnarray}
		
		Let us now compare this formulation with  the  equal-time  QKE approach of  \cite{zh1,masscorr}.  There, the energy averages  $\int dp_0 D_\mu^{(\ssE \ssM) }{\cal C}_a(x.p)= (D_t ^{(\ssE \ssM )},{\bm D}^{(\ssE \ssM)} ){\cal C}_a(x,\bm p )$   are given in terms of the electromagnetic fields  $\bm E,$ $\bm B$ as $D_t ^{(\ssE \ssM )}=\partial_t +\bm E \cdot \bm \nabla_p ,$ and $\bm D^{(\ssE \ssM) }=\bm \nabla +\bm B \times \bm \nabla_p .$  We set  the electric charge $Q=1.$ Observe that  one obtains $D_t$ given in \eqref{Dt} from $D_t ^{(\ssE \ssM )}$  by the substitution  ${\bm E}\rightarrow \bm p\cross\bm{\omega}.$  However, $\bm D^{(\ssE \ssM) }$ is quite different from  \eqref{DD}. First of all, although    $\bm D^{(\ssE \ssM) }$ is independent of $\hbar,$ in   \eqref{DD} there exists a term which is at the order of $\hbar.$  Also the $\hbar$ independent terms are not similar. Only $\bm D^{(\ssE \ssM) } $  corresponds  to  $ \bm{\partial}^{(0)}$  
	 by ${\bm B}\rightarrow\kappa E_p \bm{\omega}.$ Thus, one cannot generate our results from the ones given in \cite{masscorr}, by substituting $ \bm E, {\bm B} $ with $ \bm p\cross\bm{\omega}, \kappa E_p \bm{\omega}.$ 
		
The  transport equations at  the  first order in $\hbar$ can be read from  \eqref{8a}-\eqref{8h} as
		\begin{subequations}
			\begin{align}
				&	D_t f_0^{(0)\pm}+{\bm D}_\pm^{(0)}\cdot\mathbf{g}_1^{(0)\pm}=0,{\label{17a}}\\
				&	D_t f_1^{(0)\pm}+{\bm D}_\pm^{(0)}\cdot\mathbf{g}_0^{(0)\pm}+2mf_2^{(1)\pm}=0,{\label{17b}}\\
				&   D_tf_2^{(0)\pm}+\bm p\cdot\mathbf{g}_3^{(1)\pm}-2mf_1^{(1)\pm}=0,{\label{17c}}\\
				&	D_t f_3^{(0)\pm}-2\bm p\cdot\mathbf{g}_2^{(1)\pm}=0,{\label{17d}}\\
				&	D_t \mathbf{g}_0^{(0)\pm}+{\bm D}_\pm^{(0)} f_1^{(0)\pm}-2\bm p\cross\mathbf{g}_1^{(1)\pm}=0,{\label{17e}}\\
				&	D_t  \mathbf{g}_1^{(0)\pm}+{\bm D}_\pm^{(0)}f_0^{(0)\pm}-2\bm p\cross\mathbf{g}_0^{(1)\pm}+2m\mathbf{g}_2^{(1)\pm}=0,{\label{17f}}\\
				&	D_t \mathbf{g}_2^{(0)\pm}+{\bm D}_\pm^{(0)}\cross\mathbf{g}_3^{(0)\pm}+2\bm pf_3^{(1)\pm}-2m\mathbf{g}_1^{(1)\pm}=0,{\label{17g}}\\
				&	D_t \mathbf{g}_3^{(0)\pm}-{\bm D}_\pm^{(0)}\cross\mathbf{g}_2^{(0)\pm}-2\bm pf_2^{(1)\pm}=0.{\label{17h}}
			\end{align}
		\end{subequations}
		By plugging  \eqref{Continuous expansion} into \eqref{9a}-\eqref{9h} and employing the definition  \eqref{Continuous Shift},   the constraint equations at the first order in $\hbar$ are acquired as
		\begin{subequations}
			\begin{align}
				&	\pm E_p f_0^{(1)\pm}+ \Delta E_{f_0}^\pm-\bm p\cdot \mathbf{g}_{1}^{(1)\pm} -mf_3^{(1)\pm}=0,\label{13a}\\
				&	\pm E_p f_1^{(1)\pm}+ \Delta E_{f_1}^\pm-\bm p\cdot \mathbf{g}_{0}^{(1)\pm}=0,\label{13b}\\
				&	\pm E_p f_2^{(1)\pm}+ \Delta E_{f_2}^\pm+\frac{1}{2}{\bm D}_{\pm}^{(0)}\cdot \mathbf{g}_{3}^{(0)\pm}=0,\label{13c}\\
				&	\pm E_p f_3^{(1)\pm}+ \Delta E_{f_3}^\pm-\frac{1}{2}{\bm D}_\pm^{(0)} \cdot \mathbf{g}_{2}^{(0)\pm}-mf_0^{(1)\pm}=0,\label{13d}\\
				&	\pm E_p \mathbf{g}_0^{(1)\pm}+\Delta{\bm E}_{\mathbf{g}_0}^\pm-\bm p f_1^{(1)\pm}-\frac{1}{2} {\bm D}_{\pm }^{(0)} \cross \mathbf{g}_{1}^{(0)\pm}-m \mathbf{g}_3^{(1)\pm}=0,\label{13e}\\
				&	\pm E_p \mathbf{g}_1^{(1)\pm }+\Delta{\bm E}_{\mathbf{g}_1}^\pm-\bm pf_0^{(1)\pm}-\frac{1}{2} {\bm D}_{\pm}^{(0)} \cross \mathbf{g}_{0}^{(0)\pm}=0,\label{13f}\\
				&	\pm E_p  \mathbf{g}_2^{(1)\pm}+\Delta{\bm E}_{\mathbf{g}_2}^\pm-\bm p\cross\mathbf{g}_3^{(1)\pm}+\frac{1}{2}{\bm D}_\pm^{(0)}f_3^{(0)\pm}=0,\label{13g}\\
				& \pm E_p \mathbf{g}_3^{(1)\pm}+\Delta{\bm E}_{\mathbf{g}_3}^\pm+\bm p\cross\mathbf{g}_2^{(1)\pm}-m\mathbf{g}_0^{(1)\pm}=0.\label{13h}
			\end{align}
		\end{subequations}
		Once we are acquainted with $\Delta E_{a}^\pm (x,\bm p),$ the constraint equations \eqref{13a}-\eqref{13h} can be solved to express  the 
		first-order components of the fields in terms of $f_0^{(0)\pm}$ and $\mathbf{g}_0^{(0)\pm}.$ Some of the shell  shifts can be acquired by making use the covariant formalism as we have presented in Appendix. The remaining  shell  shifts should be determined by using the constraint and transport equations \eqref{17a}-\eqref{13h}. In conclusion, we calculated the mass shell  shifts as
		\begin{align}
			\Delta E_{f_0}^\pm&=-\frac{\kappa}{2}\bm{\omega}\cdot\mathbf{g}_0^{(0)\pm},\label{DEP0}\\
			\Delta E_{f_1}^\pm&=\mp\frac{\kappa}{2E_p}\bm p\cdot\bm{\omega} f_0^{(0)\pm},\label{DEP1}\\
			\Delta E_{f_2}^\pm&=\frac{(1+\kappa)}{2m}(\bm p\cross\bm{\omega})\cdot\mathbf{g}_0^{(0)\pm},\label{DEP2}\\
			\Delta E_{f_3}^\pm&=\mp\frac{1}{2mE_p}(\bm p\cross\bm{\omega})\cdot(\bm p\cross\mathbf{g}_0^{(0)\pm})\mp\kappa\Bigg(\frac{E_p^2 \mathbf{g}_0^{(0)\pm}-(\bm p\cdot\mathbf{g}_0^{(0)\pm})\bm p}{2mE_p}\Bigg)\cdot\bm{\omega},\label{DEP3}\\
			\Delta{\bm E}_{\mathbf{g}_0}^\pm&=-\Bigg(\frac{\kappa}{2}\bm{\omega}+\frac{\bm{\omega} \bm p^2-\bm p(\bm{\omega}\cdot\bm p)}{2E_p^2}\Bigg) f_0^{(0)\pm},\label{VDEP0}\\
			\Delta{\bm E}_{\mathbf{g}_1}^\pm&=\mp \frac{1}{2E_p}(\bm p\cross\bm{\omega})\cross\mathbf{g}_0^{(0)\pm} \mp \frac{\kappa}{2E_p}(\bm p\cdot\mathbf{g}_0^{(0)\pm})\bm{\omega},\label{VDEP1}\\
			\Delta{\bm E}_{\mathbf{g}_2}^\pm&=\frac{m\bm p\cross\bm{\omega}}{2E_p^2}f_0^{(0)\pm},\label{VDEP2}\\
			\Delta{\bm E}_{\mathbf{g}_3}^\pm&=\mp \frac{m\kappa}{2E_p}\bm{\omega}f_0^{(0)\pm}\label{VDEP3}.
		\end{align}
		It is a curious fact that although $\bm D^{(0)}$ given in \eqref{D0} is different from its electromagnetic analog $\bm D^{(\ssE \ssM) }$, we still get the  correspondence between the shell shifts  given  in   \cite{masscorr} and the ones calculated here as in \eqref{DEP0}-\eqref{VDEP3},   by  the substitution  ${\bm E}\rightarrow \bm p\cross\bm{\omega}$ and ${\bm B} \rightarrow \kappa E_p \bm{\omega}.$ 
		
		Now, the constraint equations \eqref{13a}-\eqref{13h} are employed to determine the field components at the first order in $\hbar,$ in terms of $f_0^{(0)\pm}$ and $\mathbf{g}_0^{(0)\pm}$ as follows:  
		\begin{eqnarray}
			f_1^{(1)\pm}&=&\frac{\kappa}{2E_p^2}\bm p\cdot\bm{\omega}f_0^{(0)\pm}\pm\frac{1}{E_p}\bm p\cdot\mathbf{g}_0^{(1)\pm},\label{16a}\\
			f_2^{(1)\pm}&=&\mp\frac{(1-\kappa)}{2mE_p}(\bm p\cross\bm{\omega})\cdot\mathbf{g}_0^{(0)\pm}-\frac{1}{2m}{\bm D}_\pm\cdot\mathbf{g}_0^{(0)\pm}+\frac{1}{2mE_p^2}\bm p\cdot(\bm p\cdot\bm{\partial}_\pm^{(0)})\mathbf{g}_0^{(0)\pm},\label{16b}\\
			f_3^{(1)\pm}&=&\pm\frac{m}{E_p}f_0^{(1)\pm}\mp\frac{(\bm p\cross{\bm D}_\pm^{(0)})\cdot\mathbf{g}_0^{(0)\pm}}{2mE_p} +\frac{1}{2mE_p^2}(\bm p\cross\bm{\omega})\cdot(\bm p\cross\mathbf{g}_0^{(0)\pm})-\frac{\kappa }{2m}\bm{\omega}\cdot\mathbf{g}_0^{(0)\pm}\nonumber\\
			&&-\kappa\frac{(\bm p\cdot\mathbf{g}_0^{(0)\pm})\bm p\cdot\bm{\omega}}{2mE_p^2},\label{16c}\\
			\mathbf{g}_1^{(1)\pm } &=&\pm\frac{1}{E_p}\bm pf_0^{(1)\pm}+ \frac{1}{2E_p^2}(\bm p\cross\bm{\omega})\cross\mathbf{g}_0^{(0)\pm}+ \frac{\kappa}{2E_p^2}(\bm p\cdot\mathbf{g}_0^{(0)\pm})\bm{\omega}\pm\frac{1}{2E_p} {\bm D}_{\pm}^{(0)} \cross \mathbf{g}_{0}^{(0)\pm},\label{16d}\\
			\mathbf{g}_2^{(1)\pm}&=& \frac{1}{m} \bm p\cross\mathbf{g}_0^{(1)\pm}+\frac{\bm p(\bm p\cdot\bm{\partial}^{(0)}_\pm)f_0^{(0)\pm}}{2mE_p^2}\mp\frac{1}{2mE_p}(\bm p\cross\bm{\omega})f_0^{(0)\pm}-\frac{1}{2m}{\bm D}_\pm^{(0)}f_0^{(0)\pm},\label{16e}\\
			\mathbf{g}_3^{(1)\pm} &=& \pm \frac{E_p}{m}\mathbf{g}_0^{(1)\pm}\mp\frac{1}{mE_p} \bm p(\bm p\cdot\mathbf{g}_0^{(1)\pm})+\frac{1}{2mE_p^2}\bm p\cross(\bm p\cross\bm{\omega})f_0^{(0)\pm}
			+\frac{m\kappa}{2E_p^2}\bm{\omega}f_0^{(0)\pm}\nonumber\\
			&&\pm\frac{1}{2mE_p}\bm p\cross{\bm D}_\pm^{(0)}f_0^{(0)\pm}.\label{16f}
		\end{eqnarray}
		We determined all of the 3D field components in terms of $f_0$ and $\mathbf{g}_0$ up to the first order in $\hbar.$ In the next section we  will derive their semiclassical  kinetic equations.
		
				\section{Semiclassical kinetic equations of $f_0$ and ${\mathbf{g}}_0$ }
		\label{skefg}
			Kinetic equation of the particle number density $f_0$ at the zeroth order in $\hbar$  can be  easily derived from  \eqref{12d} and \eqref{17a} as
		\begin{equation}\label{f0trans}
			\Bigg(D_t\pm\frac{\bm p}{E_p}\cdot\bm{\partial}^{(0)}_\pm\Bigg)f_0^{(0)\pm}=0.
		\end{equation}
		By employing \eqref{17b} and \eqref{17h} we can get  kinetic equation of the spin density ${\mathbf{g}}_0$ at the zeroth order in $\hbar$ as follows:
		\begin{equation}\label{g0trans}
			\Bigg(D_t\pm\frac{\bm p}{E_p}\cdot\bm{\partial}_\pm^{(0)}\Bigg)\mathbf{g}_0^{(0)\pm}    =\frac{1}{E_p^2}(\bm p\cross\bm{\omega})(\bm p\cdot\mathbf{g}_0^{(0)\pm})-\kappa \bm{\omega}\cross\mathbf{g}_0^{(0)\pm}.
		\end{equation}
		
		Let us now  derive the kinetic equations of  $f_0$ and $\mathbf{g}_0$ at next-to-leading order in $\hbar.$ To carry out our calculations 
	the transport equations  at the second order in $\hbar$ are needed. They can be acquired by making use of  \eqref{DD} in  \eqref{8a}-\eqref{8h}:
		\begin{subequations}
			\begin{align}
				&	D_t f_0^{(1)\pm}+{\bm D}_\pm^{(0)} \cdot \mathbf{g}_1^{(1)\pm}+\frac{\kappa\bm{\omega}}{2E_p}\bm p\cdot(\bm{\omega}\cross\bm{\nabla}_p)\mathbf{g}_0^{(0)\pm}=0, \label{8asecond}\\
				&	D_t f_1^{(1)\pm}+{\bm D}_\pm^{(0)}\cdot\mathbf{g}_0^{(1)\pm}+\frac{\kappa(\bm{\omega}\cdot\bm p)}{2E_p^2}\bm p\cdot(\bm{\omega}\cross\bm{\nabla}_p)f_0^{(0)\pm}+2mf_2^{(2)\pm}=0, \label{8bsecond}\\
				&	D_t f_2^{(1)\pm} + 2\bm p\cdot\mathbf{g}_3^{(2)\pm}-2mf_1^{(2)\pm}=0, \label{8csecond}\\
				&	D_t f_3^{(1)\pm} - 2\bm p\cdot\mathbf{g}_2^{(2)\pm}=0,\label{8dsecond}\\
				&	D_t \mathbf{g}_0^{(1)\pm}+{\bm D}_\pm^{(0)} f_1^{(1)\pm}\pm\frac{\kappa^2(\bm{\omega}\cdot\bm p)}{2E_p}\bm{\omega}\cross\left(\frac{\bm p}{E_p^2}-\bm{\nabla}_p\right)f_0^{(0)\pm}-2\bm p\cross\mathbf{g}_1^{(2)\pm}=0, \label{8esecond}\\
				&	D_t \mathbf{g}_1^{(1)\pm}+{\bm D}_\pm^{(0)} f_0^{(1)\pm} -\frac{\kappa^2}{2}(\bm{\omega}\cross\bm{\nabla}_p)(\bm{\omega}\cdot\mathbf{g}_0^{(0)\pm})-2\bm p\cross\mathbf{g}_0^{(2)\pm}+2m\mathbf{g}_2^{(2)\pm}=0, \label{8fsecond}\\
				&	D_t \mathbf{g}_2^{(1)\pm} + {\bm D}_\pm^{(0)} \cross \mathbf{g}_3^{(1)\pm} \mp\frac{m\kappa^2}{2E_p}\bm{\omega}\cross\left(\bm{\omega}\cross\left(\frac{\bm p}{E_p^2}-\bm{\nabla}_p\right)\right)f_0^{(0)\pm}+2\bm pf_3^{(2)\pm}-2m\mathbf{g}_1^{(2)\pm}=0, \label{8gsecond}\\
				&	D_t \mathbf{g}_3^{(1)\pm}-{\bm D}_\pm^{(0)} \cross \mathbf{g}_2^{(1)\pm} +\frac{m\kappa\bm{\omega}}{2E_p^2} \bm p\cdot(\bm{\omega}\cross\bm{\nabla}_p)f_0^{(0)\pm}-2\bm pf_2^{(2)\pm}=0. \label{8hsecond}
			\end{align}
		\end{subequations}
We would like to emphasize the fact  that at this order the resemblance  between $\bm D^{(\ssE \ssM) }$ and \eqref{DD} is completely lost due to the presence of the $\hbar$-order term in the latter.
		
		By  employing  \eqref{VDEP1}, \eqref{16d} and \eqref{8asecond},  we derived  the time evolution of $f_0^{(1)\pm}$ in terms of   $\mathbf{g}_0^{(0)\pm}$ as 
		\begin{equation}\label{transf01}
			\begin{aligned}
				\Bigg (D_t \pm\frac{\bm p}{E_p}\cdot \bm{\partial}_\pm^{(0)}\Bigg) f_0^{(1)\pm}=
				& -\frac{\kappa}{2E_p^2}(\bm{\omega}\cross\bm p)\cdot(\bm{\partial}_\pm^{(0)}\cross\mathbf{g}_0^{(0)\pm})   \\
				&+ \frac{1}{2E_p^2}(\bm p\cross\bm{\omega})  \cdot\left( \bm{\nabla}\cross\mathbf{g}_0^{(0)\pm}\right)
				- \frac{\kappa}{2E_p^2}\bm{\omega}\cdot(\bm p\cdot\bm{\nabla})\mathbf{g}_0^{(0)\pm}.
			\end{aligned}
		\end{equation}
		After some cumbersome calculations by making use of \eqref{16a}, \eqref{16e}, \eqref{16f}, \eqref{f0trans}, \eqref{8esecond} 
		and \eqref{8gsecond}, we obtained  the dynamical evolution of $\mathbf{g}_0^{(1)\pm}$ depending  on $f_0^{(0)\pm}$ as
		\begin{equation}\label{g_0^1 trans}
			\begin{aligned}
				\Bigg (D_t\pm\frac{\bm p}{E_p}\cdot\bm{\partial}_\pm^{(0)}\Bigg)\mathbf{g}_0^{(1)\pm}=&- \kappa(\bm{\omega}\cross\mathbf{g}_0^{(1)\pm})+ \frac{\bm p\cdot\mathbf{g}_0^{(1)\pm}}{E_p^2}(\bm p\cross\bm{\omega})\\
				&-\frac{\kappa \bm{\omega}}{2E_p^2} \bm p\cdot\bm{\nabla}f_0^{(0)\pm}    +\frac{\bm p(\bm p\cdot\bm{\omega})}{2E_p^4}(\bm p\cdot\bm{\partial}_\pm^{(0)})f_0^{(0)\pm}\\
				&-\frac{\bm p^2(\bm p\cdot\bm{\omega})}{2 E_p^4}{\bm D}_\pm^{(0)}f_0^{(0)\pm}\mp\frac{\bm p\cdot\bm{\omega}}{2E_p^3}(\bm p\cross\bm{\omega})f_0^{(0)\pm}\\
				&\pm\frac{\kappa}{2E_p}\bm p\cross \Bigg(-\bm{\omega}^2  \bm{\nabla}_p +  \bm{\omega}
				(\bm{\omega}\cdot\bm{\nabla}_p)+\frac{1}{E_p^2}\bm{\omega}(\bm p\cdot\bm{\omega})\Bigg)f_0^{(0)\pm}\\
				&\mp\frac{1}{2E_p^3}\bm p\cross(\bm p\cross\bm{\omega})D_t f_0^{(0)\pm}.
			\end{aligned}
		\end{equation}
		We established the semiclassical kinetic equations of $f_0$ and $\mathbf{g}_0$. It is also possible to deal with the kinetic equations of some other components of the 3D Wigner function like $f_1$ and $\mathbf{g}_3,$ where the latter is related to the magnetic dipole moment.
		
	\section{	Kinetic Theories of   the right-  and left-handed fermions}
		\label{masKT}
		
		In  heavy-ion collisions, because of  considering the constituent  quarks of the quark-gluon plasma as massless,  one expects  that the collective dynamics yield  the chiral vortical and separation  effects due to vorticity.  We would like to study how  the quark  mass   affects  this picture. To study the mass corrections, we need  the kinetic equations satisfied by the right- and left-handed distribution functions  $f_\ssR,  f_\ssL,$ defined by 
		\begin{equation}
			f_\chi =	\frac{1}{2}(f_0  +\chi  f_1 ),
		\end{equation}
		where $\chi=\{+,-\}$, and  $f_+ \equiv f_\ssR$ and $f_- \equiv  f_\ssL$.  However, the 3D kinetic equations \eqref{f0trans}, \eqref{g0trans} and \eqref{transf01}, \eqref{g_0^1 trans} are given in terms of $f_0$ and $\mathbf{g}_0.$   Thus, we have to specify the spin current $\mathbf{g}_0$, by respecting the relations \eqref{12a} and \eqref{16a}. 
		First, let the direction of the spin current be parallel to $\bm p.$  Then, \eqref{12a} implies that
		\begin{equation}\label{g00f10}
			\mathbf{g}_0^{(0)\pm} = \pm \frac{E_p }{\bm p^2} \bm pf_1^{(0)\pm}.
		\end{equation}
		By plugging \eqref{g00f10} into the classical   kinetic equation of the spin current given by \eqref{g0trans}, we find
		\begin{equation}
			\Bigg(D_t  \pm\frac{\bm p}{E_p}\cdot\bm{\partial}_\pm^{(0)}\Bigg)f_1^{(0)\pm}   = 0 .
		\end{equation}
		Recall that it has the same form with the classical kinetic equation satisfied by $f_0^{(0)\pm}$, \eqref{f0trans}.
		Additionally, \eqref{g00f10} allows us to write the right-hand side of  \eqref{transf01}  in terms of $f_1^{(0)\pm}:$ 
		\begin{equation}\label{f0f1trans}
			\begin{aligned}
				\Bigg (D_t \pm\frac{\bm p}{E_p}\cdot \bm{\partial}_\pm^{(0)}\Bigg) f_0^{(1)\pm}=&\pm\frac{1}{2E_p\bm p^2}\Bigg({(1+\kappa)}(\bm p\cross(\bm p\cross\bm{\omega}))-{\kappa}(\bm{\omega}\cdot\bm p)\bm p\Bigg)\cdot\bm{\nabla}f_1^{(0)\pm}
				\\
				&+\frac{\kappa^2}{2\bm p^2}(\bm p\cdot\bm{\omega})(\bm p\cross\bm{\omega})\cdot\bm{\nabla}_p f_1^{(0)\pm}.
			\end{aligned}
		\end{equation}
		Now, we desire to find the kinetic equation satisfied by $f_1^{(1)\pm}$. For this purpose, first observe that Eq. \eqref{16a} can be solved as
		\begin{equation}\label{g_0^1 sol}
			\mathbf{g}_0^{(1)\pm} = \pm\frac{E_p\bm p}{\bm p^2}f_1^{(1)\pm}\mp\frac{\kappa\bm{\omega}}{2E_p}f_0^{(0)\pm}\pm E_p\bm p\cross{\bm F}^\pm,
		\end{equation}
		where ${\bm F}^\pm$ is a free vector field which will be fixed shortly. Then, by plugging \eqref{g_0^1 sol} into \eqref{g_0^1 trans} and then multiplying it with 
		$\pm \bm p / E_p,$ we find
		\begin{equation}\label{f_1^1 trans}
			\begin{aligned}
				\Bigg (D_t\pm\frac{\bm p}{E_p}\cdot\bm{\partial}_\pm^{(0)}\Bigg)f_1^{(1)\pm}=&-(\bm p\cross(\bm p\cross\bm{\omega}))\cdot{\bm F}^\pm
				\mp\frac{\kappa\bm p\cdot\bm{\omega}}{2E_p^3} \bm p\cdot\bm{\nabla}f_0^{(0)\pm}\\&
				+\frac{\kappa\bm{\omega}}{2E_p^2}\cdot\bm p\Bigg (D_t\pm\frac{\bm p}{E_p}\cdot\bm{\partial}_\pm^{(0)}\Bigg)f_0^{(0)\pm}.
			\end{aligned}
		\end{equation}
		To have an equation compatible with \eqref{f0f1trans}, we choose ${\bm F}^\pm$ to be
		\begin{equation}
			{\bm F}^\pm = \mp\frac{(\kappa+1)}{2E_p\bm p^2}\bm{\nabla}f_0^{(0)\pm}-\frac{\kappa^2}{2\bm p^2}\bm{\omega}\cross\bm{\nabla}_p f_0^{(0)\pm}.
		\end{equation}
		By inserting it into \eqref{f_1^1 trans} one gets the kinetic equation
		\begin{equation}\label{f_1^1 transtrans}
			\begin{aligned}
				\Bigg (D_t\pm\frac{\bm p}{E_p}\cdot\bm{\partial}_\pm^{(0)}\Bigg)f_1^{(1)\pm}=&\pm\frac{(\kappa+1)}{2E_p\bm p^2}(\bm p\cross(\bm p\cross\bm{\omega}))\cdot\bm{\nabla}f_0^{(0)\pm}+\frac{(\bm p\cdot\bm{\omega})\kappa^2}{2\bm p^2}(\bm p\cross\bm{\omega})\cdot\bm{\nabla}_p f_0^{(0)\pm}\\&\mp\frac{\kappa\bm{\omega}\cdot\bm p}{2E_p^3} \bm p\cdot\bm{\nabla}f_0^{(0)\pm}+\frac{\kappa\bm{\omega}\cdot\bm p}{2E_p^2}\Bigg (D_t\pm\frac{\bm p}{E_p}\cdot\bm{\partial}_\pm^{(0)}\Bigg)f_0^{(0)\pm}.
			\end{aligned}
		\end{equation}
		The last term can be set equal to zero due to (\ref{f0trans}). However, instead of doing that, 
		we add a similar vanishing term to the right-hand side of \eqref{f0f1trans}:
		\begin{equation}\label{f0f1transtrans}
			\begin{aligned}
				\Bigg (D_t \pm\frac{\bm p}{E_p}\cdot \bm{\partial}_\pm^{(0)}\Bigg) f_0^{(1)\pm}=&    \pm\frac{(\kappa+1)}{2E_p \bm p^2}(\bm p\cross(\bm p\cross\bm{\omega}))\cdot\bm{\nabla}f_1^{(0)\pm}
				+\frac{(\bm p\cdot\bm{\omega})\kappa^2}{2\bm p^2}(\bm p\cross\bm{\omega})\cdot\bm{\nabla}_pf_1^{(0)\pm}
				\\&\mp \frac{\kappa}{2E_p} \frac{\bm{\omega}\cdot\bm p}{\bm p^2}(\bm p\cdot\bm{\nabla})f_1^{(0)\pm}+\frac{\kappa\bm{\omega}\cdot\bm p}{2E_p^2}\Bigg (D_t\pm\frac{\bm p}{E_p}\cdot\bm{\partial}_\pm^{(0)}\Bigg)f_1^{(0)\pm}.
			\end{aligned}
		\end{equation}
		Notice that adding the last term 
		is equivalent to a shift of $f_0^{(1)\pm}$ with the term $\frac{\kappa \bm{\omega}\cdot \bm p}{2E_p^2}f_1^{(0)\pm}.$ 
		By combining \eqref{f_1^1 transtrans} and \eqref{f0f1transtrans}, we find the kinetic equations
		\begin{equation}\label{xitransport equation}
			\begin{aligned}
				&\Bigg\{\Bigg(1-\hbar\frac{\chi\kappa(\bm p\cdot\bm{\omega})}{2E_p^2}\Bigg)\partial_t+\Bigg[\Bigg(1-\hbar\frac{\chi\kappa(\bm p\cdot\bm{\omega})}{2E_p^2}\Bigg)(\kappa+1)-\hbar\frac{\chi(\bm p\cdot\bm{\omega})\kappa^2}{2\bm p^2}\Bigg](\bm p\cross\bm{\omega})\cdot\bm{\nabla}_p
				\\&+\Bigg[ \pm\frac{\bm p}{E_p}\mp\hbar\frac{\chi(\kappa+1)}{2E_p \bm p^2}\bm p\cross(\bm p\cross\bm{\omega})\pm \hbar\frac{\chi\kappa m^2 (\bm p\cdot\bm{\omega})}{4E_p^3\bm p^2}\bm p \Bigg]\cdot \bm{\nabla}\Bigg\}f_\chi^\pm\\
				&=\pm\hbar\frac{\chi\kappa m^2\bm p\cdot\bm{\omega}}{4E_p^3 \bm p^2} \bm p\cdot\bm{\nabla}f_{-\chi}^\pm.
			\end{aligned}
		\end{equation}
		Therefore, we establish the kinetic theory 
		\begin{equation}
			\Bigg[\sqrt{\eta}_\chi^\pm
		\partial_t
			+(\sqrt{\eta}\Dot{\mathbf{x}})^\pm_\chi\cdot\bm{\nabla}+(\sqrt{\eta}\Dot{\bm p})_\chi^\pm\cdot\bm{\nabla}_p\Bigg]f_{\chi}^\pm=\pm \hbar\frac{\chi m^2 \bm p\cdot\bm{\omega}}{4E_p^3 \bm p^2}\bm p\cdot\bm{\nabla}f_{-\chi}^\pm, \label{BeM}
		\end{equation}
		with
		\begin{align}
			\sqrt{\eta}_\chi^\pm&=1-\hbar\frac{\chi(\bm p\cdot\bm{\omega})}{2E_p^2}, \label{kt1}\\
			(\sqrt{\eta}\Dot{\mathbf{x}})^\pm_\chi &= \pm\frac{\bm p}{E_p}\pm\hbar\frac{\chi\bm{\omega}}{E_p}\mp\hbar\frac{\chi(\bm p\cdot\bm{\omega})\bm p}{4E_p}\Bigg( \frac{3}{\bm p^2}+\frac{1}{E_p^2}\Bigg),   \label{kt2}\\
			(\sqrt{\eta}\Dot{\bm p})_\chi^\pm&=\Bigg[2\Bigg(1-\hbar\frac{\chi(\bm p\cdot\bm{\omega})}{2E_p^2}\Bigg)-\hbar\frac{\chi(\bm p\cdot\bm{\omega})}{2\bm p^2}\Bigg](\bm p\cross\bm{\omega}).   \label{kt3}
		\end{align}
		We set  $\kappa=1$ for acquiring  the Coriolis force correctly.   The term appearing on the right-hand side of \eqref{BeM} shows that for the massive fermions right- and left-handed distributions cannot be decoupled.
	
		Getting inspiration from  the left- and right-handed decompositions of the distribution functions, $f_0=f_\ssR +f_\ssL,\ f_1=f_\ssR-f_\ssL,$ we write the shell shifts in \eqref{DEP0} and \eqref{DEP1} as
		\begin{eqnarray}
			\Delta E_{f_0}^\pm =\Delta E_{f_0\ssR}^\pm+\Delta E_{f_0\ssL}^\pm&= &\mp\frac{E_p\, \bm p\cdot\bm{\omega}}{2\bm p^2}(f_R^{(0)\pm}-f_L^{(0)\pm}),\\
			\Delta E_{f_1}^\pm=	\Delta E_{f_1\ssR}^\pm-\Delta E_{f_1\ssL}^\pm &= &\mp\frac{\bm p\cdot\bm{\omega}}{2E_p}(f_R^{(0)\pm}+f_L^{(0)\pm}).
		\end{eqnarray}
		Hence,  for the left- and right-handed fermions we define
		\begin{equation}
			\Delta E_\chi^\pm = \mp\frac{\chi}{4E_p}\Bigg(1+\frac{E_p^2}{\bm p^2}\Bigg)\bm p\cdot\bm{\omega}.
		\end{equation}
		Therefore, the  dispersion relations are
		\begin{equation}
			\label{disper}
			\epsilon_{p,\chi}^{\pm}=\pm E_p \mp\hbar\frac{\chi}{4E_p}\Bigg(1+\frac{E_p^2}{\bm p^2}\Bigg)\bm p\cdot\bm{\omega}.
		\end{equation}
		
		The particle number current density can be written in terms of the equilibrium distribution function as
		\begin{equation}
			\label{curden}
			\bm j_\chi^\pm =\int \frac{d^3 \bm p}{(2\pi)^3}    (\sqrt{\eta}\Dot{\mathbf{x}})^{\pm}_\chi f_\chi^{eq\pm}(\epsilon_{p,\chi}^\pm).
		\end{equation}
		Let the equilibrium distribution function be taken as the Fermi-Dirac distribution:
		\begin{equation}
			\label{FD}
			f_\chi^{eq\pm}(\epsilon_{p,\chi}^\pm)=\frac{1}{e^{\pm(\epsilon_{p,\chi}^\pm-\mu_\chi)/T}+1},
		\end{equation}
		where $\mu_\chi$ is the chiral chemical potential, $T$ is the temperature, and we employed the dispersion relations \eqref{disper}. 
		We can expand  \eqref{FD} in Taylor series as
		\begin{equation}
				f_\chi^{eq\pm}(\epsilon_{p,\chi}^\pm) \approx  f_\chi^{eq\pm}(E_p)\mp\hbar\frac{\chi}{4E_p}\Bigg(1+\frac{E_p^2}{\bm p^2}\Bigg)\bm p\cdot\bm{\omega} \frac{df_\chi^{eq\pm}(E_p)}{dE_p},
		\end{equation}
		where
		\begin{equation}
			f_\chi^{eq\pm}(E_p)=\frac{1}{e^{(E_p\mp\mu_\chi)/T}+1}.
		\end{equation}
		Notice that the equilibrium distribution function  depends only on the magnitude of the momentum. Therefore, we can evaluate the angular part of the integral in  \eqref{curden}, yielding
		\begin{equation}\label{j when f expanded}
			\begin{aligned}
				\bm j_\chi^\pm &= \hbar\chi\bm{\omega}\int \frac{d |\bm p|}{24\pi^2}\bm p^2 \Bigg[\Bigg(\pm\frac{8}{E_p} \pm \frac{ m^2 }{E_p^3}\Bigg) f_\chi^{eq\pm}(E_p)      -\Bigg(\frac{1}{E_p^2}+\frac{1}{\bm p^2}\Bigg)\bm p^2 \frac{d f_\chi^{eq\pm}(E_p)}{dE_p}\Bigg] .    
			\end{aligned}
		\end{equation}
		Since the classical terms vanish,  the current densities are at the order of $\hbar.$ Then, the  vector and axial vector current densities, $\bm j_\ssV = \bm j_\ssR+\bm j_\ssL , \ \ \bm j_\ssA =\bm j_\ssR-    \bm j_\ssL,$ are accomplished as
		\begin{eqnarray}
			\bm j_{\ssV,\ssA}& = &\sum_\pm\hbar\bm{\omega}\int \frac{d |\bm p|}{24\pi^2}\bm p^2 \Bigg[\Bigg(\pm\frac{8}{E_p} \pm \frac{ m^2 }{E_p^3}\Bigg) f_{\ssV,\ssA}^{eq\pm}(E_p)      -\Bigg(\frac{1}{E_p^2}+\frac{1}{\bm p^2}\Bigg)\bm p^2 \frac{d f_{\ssV,\ssA}^{eq\pm}(E_p)}{dE_p}\Bigg] \nonumber \\
			&\equiv & \sigma_{\ssV, \ssA}\bm{\omega}.
		\end{eqnarray}
		We introduced
		\begin{equation}
			f_{\ssV,\ssA}^\pm =\frac{1}{e^{(E_p\mp \mu_R)/T}+1}\mp\frac{1}{e^{(E_p\mp \mu_L)/T}+1}.
		\end{equation}
		
		Observe that at zero temperature, the distribution functions transform into the Heaviside step function for positive energy particles and vanish for negative energy particles.
		For simplicity, let us set $\mu_\ssR=\mu_\ssL\equiv\mu$. Then, the vector current  vanishes and the axial vector current gives  
		\begin{equation}
			\lim_{T\rightarrow 0}\sigma_\ssA =\frac{\hbar}{2\pi^2}\Bigg[\frac{3\mu^2-m^2}{3\mu}\sqrt{\mu^2-m^2}-\frac{1}{2}m^2\ln (\frac{\mu+\sqrt{\mu^2-m^2}}{m})\Bigg] \theta(\mu-m).
		\end{equation}
		In  the small mass  limit we get
		\begin{equation}
			\label{cvem}
			\lim_{T\rightarrow 0}\sigma^+_A =\frac{\hbar}{2\pi^2} \left[\mu\sqrt{\mu^2-m^2}-\frac{m^2}{3}\right]\theta(\mu-m).
		\end{equation}
		This result is in harmony with the field theoretic calculations performed by means of the Kubo  formula in  \cite{bgb-m,lin-yang} when one ignores the last term.  However, if one excludes  the  chemical potential terms only the $m^2$ term survives. In fact,  the latter  is similar to  the small  mass correction  obtained  in curved space  in		\cite{FlFu}.
		
		Although  we do not consider the electromagnetic fields,  by inspecting the kinetic equations obtained in \cite{masscorr} one can observe that the time evolution of spatial coordinates linear in the magnetic field can be acquired from \eqref{kt2} by substituting $\bm \omega$ with $\bm B /E_p.$ Hence, the related axial  current will produce the finite mass corrections to the kinetic coefficient of the  chiral separation effect.
		
		Let us inspect the chiral (massless) limit:  First of all, \eqref{BeM}-\eqref{kt3} generate  the chiral kinetic theory  
		\begin{equation}
			\Bigg[\sqrt{\eta}_\chi^{\ssC \pm}
		\partial_t
			+(\sqrt{\eta}\Dot{\mathbf{x}})^{\ssC\pm}_\chi \cdot\bm{\nabla}+(\sqrt{\eta}\Dot{\bm p})_\chi^{\ssC\pm}  \cdot\bm{\nabla}_p\Bigg]f_{\chi}^\pm=0 , \label{BeM0}
		\end{equation}
		with
		\begin{align}
			\sqrt{\eta}_\chi^{\ssC \pm}&=1-\hbar\frac{\chi\bm{\omega}\cdot\bm p}{2\bm p^2}, \label{cv1}\\
			(\sqrt{\eta}\Dot{\mathbf{x}})^{\ssC \pm}_\chi&=\pm\frac{\bm p}{|{\bm p}|}\mp\hbar\frac{\chi}{|{\bm p}|^3}\bm p(\bm p\cdot\bm{\omega})
			\pm\hbar\frac{\chi}{|{\bm p}|}\bm{\omega}, \label{cv2}\\
			(\sqrt{\eta}\Dot{\bm p})^{\ssC \pm}_\chi&=2\bm p\cross\bm{\omega}-\hbar\chi\frac{3(\bm p\cdot\bm{\omega})}{2\bm p^2}(\bm p\cross\bm{\omega}). \label{cv3}
		\end{align}
		Then,  \eqref{disper} gives the dispersion relation for chiral particles as
		\begin{equation}
			\epsilon_{p,\chi}^{\ssC\pm}=\pm|{\bm p}| \mp\hbar\frac{\chi}{2}\hat{\bm p}\cdot\bm{\omega}.
		\end{equation}
		It is consistent with the  dispersion relation obtained in  \cite{cssyy,dky, lgmh}. Moreover, the dynamical evolution of the spatial coordinate vector, \eqref{cv2}, coincides with the one established in \cite{dk}. 
		Let the equilibrium distribution be given by the Fermi-Dirac distribution:
		\begin{equation}
			\label{FDC}
			f_\chi^{eq\pm}(\epsilon_{p,\chi}^{\ssC \pm})=\frac{1}{e^{\pm(\epsilon_{p,\chi}^{\ssC\pm}-\mu_\chi)/T}+1},
		\end{equation}
		Thus, the chiral particle number current densities are acquired as
		\begin{equation}
			\bm j_\chi^{\ssC\pm} = \hbar\chi\bm{\omega}\int \frac{d |\bm p|}{3\pi^2} \Bigg(\pm{ |\bm p|} f_\chi^{eq\pm}( |\bm p|)-\frac{1}{4}\bm p^2 
			\frac{df_\chi^{eq\pm}( |\bm p|)}{d|\bm p|}\Bigg).
		\end{equation}
		We can perform the integrals and obtain  the vector and axial vector currents  as
		$$
		\bm j_\ssV =\hbar \frac{\mu\mu_\ssA}{2\pi^2} \bm \omega,\ \   \bm j_\ssA =\hbar \left(\frac{T^2}{12}+\frac{\mu^2 +\mu^2_\ssA}{4\pi^2} \right)\bm \omega,
		$$
		where $\mu=\mu_\ssR+\mu_\ssL,$ $\mu_\ssA=\mu_\ssR-\mu_\ssL .$
		These coincide with the results reported in  \cite{glpww}. Therefore, we conclude that in the massless limit  the chiral vector and separation effects are generated correctly.
		
		\section{Discussions}
		\label{disc}
		
		The VQKE of the Wigner function leads  to the transport  equations of  the  components of the covariant  Wigner function. 
		By integrating them over 	$p_0,$ we write the equations which the components of 3D Wigner function obey. They can be separated into  the transport and constraint  equations. The vector component of the covariant $D_\mu=(D_t,\bm D)$ operator  depends explicitly on  $p_0$ as in \eqref{Dpo}.  Hence, also the transport equations depend explicitly on  $p_0,$ in contrast to the transport equations  which  have been defined in \cite{zh1,masscorr}.   $p_0$ integrals are performed by employing the on-shell conditions of  the covariant fields. Then, $\bm D$ effectively becomes as in  \eqref{DD}, which is very different from the $\bm D^{(\ssE \ssM)}$  appearing in \cite{zh1,masscorr}. Therefore, it is not possible to generalize the method of \cite{masscorr} directly to our case. Nevertheless, to study the 3D transport and constraint equations,  we follow the method proposed in \cite{masscorr} and  let each component of the Wigner function satisfy a different on-shell condition at   $\hbar$ order.
		We presented these  shell shifts and by plugging them into the constraint equations we expressed the components of the 3D Wigner function at  first order in terms of  $f_0, \mathbf{g}_0.$    We consider $f_0$ and $\mathbf{g}_0$ as  independent components. The main objective is to establish the semiclassical kinetic equations of the fields which are chosen  as the independent set of  components.  After some cumbersome calculations we acquired them as in \eqref{f0trans}, \eqref{g0trans} and \eqref{transf01}, \eqref{g_0^1 trans}.
		
		To accomplish the mass corrections to the chiral (massless) kinetic equations, we have fixed the spin current   $\mathbf{g}_0$ in terms of $f_0$ and $f_1.$ Then, we derived the  kinetic equations of right- and left-handed distribution functions  in \eqref{xitransport equation}, which provide us the kinetic theories of the right- and left-handed fermions. We acquired their dispersion relations and calculated particle number current densities by choosing the equilibrium distribution functions   appropriately.   We have shown that the massless case generates the chiral vortical and separation effects correctly. Therefore, we  succeeded in accomplishing the mass corrections to the chiral effects.
		
		In principle we can consider  a system with the nonvanishing  linear acceleration $a_\mu=u_\nu \partial^{\nu}u_\mu,$ by adding the term $(a_\mu u_\nu-a_\nu u_\mu)$ to $w_{\mu \nu}$ given in  \eqref{w_munu}.  
		The procedure which we employed here in  obtaining mass shell shifts, relies on  the solutions of the  covariant equations reported in \cite{dk-m}. 
		Hence, to deal with  nonvanishing  $a_\mu ,$   one should first study  solutions of the kinetic equations obeyed by the covariant Wigner function components with this altered $w_{\mu \nu},$  which would be complicated.
		
		A challenging future research direction is the study of 3D transport theory of  VQKE in the presence of  electromagnetic fields. As far as the contributions linear in electromagnetic fields and vorticity are concerned, this can simply be achieved by gathering the results obtained here and the ones reported in \cite{masscorr}, as we discussed after \eqref{cvem}. However, establishing kinetic equations of $f_0$ and $\mathbf{g}_0$ up to the first order in $\hbar$ in the  presence of only vorticity or electromagnetic fields is already  very difficult. Thus, when they are considered together, deriving the semiclassical  kinetic equations of  $f_0$ and $\mathbf{g}_0$  will be a demanding task. Covariant kinetic equations established in \cite{dk-m} may give some hints to solve this problem.
		
		Kinetic equations are useful mainly when collisions are taken into account. Thus, incorporating scatterings in the 3D formulation  is desired.  Unfortunately, we do not know how to do it for the VQKE.  In principle, this can be achieved by considering  the collisions in the covariant approach first and then deal with the 3D VQKE by integrating them over $p_0.$ This will generate collision terms on the right-hand side of \eqref{8a}-\eqref{9h}.  In this respect, the methods employed in \cite{yhh-col1,swsrw-col} can be useful. The other method would be to introduce  collisions to the kinetic equations of the independent set of fields \eqref{f0trans}, \eqref{g0trans}, \eqref{transf01}, \eqref{g_0^1 trans}. This is another challenging open problem.
		
		\renewcommand{\appendixname}{APPENDIX}
		\appendix*
		
		\setcounter{equation}{0}
		
		\section{CALCULATION OF SHELL SHIFTS FROM THE COVARIANT FORMULATION}
		\label{appendix}
	
		Semiclassical solutions of the  kinetic equations obeyed by  the components of 4D Wigner function,   \eqref{real1}-\eqref{imag5},  have been presented in \cite{dk-m}. By inspecting $\hbar$-order components of those solutions, one observes that some of them are expressed  in  the form
		\begin{equation}
			\label{Appendix eq 2}
			{\cal C}_i^{(1)} = \beta_i^{(1)} \delta(p^2-m^2)- \Delta E_{i}(p) \delta'(p^2-m^2),
		\end{equation}
where $\beta_i^{(1)} $ are first-order fields.	One can notice that the  3D mass shell  shifts for these fields  can be obtained as
		\begin{equation}
			\Delta E_{i}(\bm p) = \int \Delta E_i(p) \delta(p^2-m^2) dp_0.
		\end{equation}
		In this fashion,  we calculated the on-shell energy shifts for the following  components of 4D Wigner function.
		\begin{itemize}
			\item 		The scalar field ${\cal F}:$  
		\begin{equation}
			{\cal F}^{(1)}=m\delta(p^2-m^2)f_V^1
			-\frac{m}{2} \delta'(p^2-m^2)f_A^0 \Sigma^{(0)}_{\mu\nu}w^{\mu\nu}.
		\end{equation} 
		$f_A^0,f_V^1$ are scalars and
		$\Sigma^{(0)}_{\mu\nu}=-(1/m)\varepsilon_{\mu\nu\alpha \beta} p^\alpha s^\beta,$ where $s_\mu$ is the spin quantization  direction four-vector.
		\begin{equation}
			\Delta E^\pm_{f_3}(\bm p) =\pm  \frac{m}{2}\int dp_0   \Sigma^{(0)}_{\mu\nu}w^{\mu\nu} f_A^0  \delta(p^2-m^2)
			= \mp\frac{1}{2E_p}(\bm p\cross\bm{\omega})\cdot\mathbf{g}_2^{(0)\pm}-\kappa\mathbf{g}_3^{(0)\pm}\cdot\bm{\omega}.
		\end{equation}
	
		\item The axial-vector field ${\cal A}_\mu :$ 
		\begin{equation}
					{\cal A}^{(1)}_\mu =  \frac{1}{2}\epsilon_{\mu\nu\rho\sigma}p^\nu\Sigma^{(1) \rho \sigma}\delta(p^2-m^2)-\frac{1}{2}\epsilon_{\mu\nu\rho\sigma}w^{\rho\sigma}p^\nu f_V^0\delta'(p^2-m^2).
		\end{equation}
		$\Sigma^{(1)}_{ \mu\nu}$ is an antisymmetric tensor field and $f_V^0$ is a scalar.  
		
		\noindent
		For ${\cal A}_0 ,$ 
		\begin{align}
			\Delta E_{f_1}^\pm(\bm p) =  \pm \frac{1}{2}\int dp_0 \epsilon_{ijk}w^{jk}p^i f_V^0\delta(p^2-m^2)  = \mp\frac{\kappa}{2E_p}\bm p\cdot\bm{\omega} f_0^{(0)\pm}.
		\end{align}
		For $\bm{{\cal  A} },$ 
		\begin{align}
			\Delta {E}^{i\pm}_{\mathbf{g}_0}(\bm p)=\mp \frac{1}{2}\int dp_0\epsilon^{i\nu\alpha\beta}w_{\alpha\beta}p_\nu f_V^0 \delta(p^2-m^2)=-\Bigg(\frac{\kappa}{2}\bm{\omega}+\frac{\bm{\omega} \bm p^2-\bm p(\bm{\omega}\cdot\bm p)}{2E_p^2}\Bigg)^i f_0^{(0)\pm}.
		\end{align}
		
		\item The antisymmetric tensor field  $S_{\mu\nu}:$
		\begin{equation}
					S^{(1)}_{\mu\nu} = m\Sigma^{(1)}_{\mu\nu} \delta(p^2-m^2) - m w_{\mu\nu} f_V^0\delta'(p^2-m^2).
		\end{equation}
		For  $S_{0i},$ 
		\begin{align}
			\Delta E_{\mathbf{g}_2}^{i\pm}(\bm p) = \pm m\int dp_0 w^{i0}f_V^0 \delta(p^2-m^2) = \frac{m(\bm p\cross\bm{\omega})^i}{2E_p^2}f_0^{(0)\pm}.
		\end{align}
	For $S_{ij},$ 
		\begin{align}
			\Delta E_{\mathbf{g}_3}^{i\pm}(\bm p)&=\pm\frac{m}{2}\int dp_0 \epsilon^{ijk}w_{jk}f_V^0 \delta(p^2-m^2) =\mp \frac{m\kappa}{2E_p}{\omega}^i f_0^{(0)\pm}.
		\end{align}
	\end{itemize}
		These  are the  mass shell  shifts  which we determine from the covariant approach.
		
		\bibliography{3d-VQKE-PRD-rev}
	
	\end{document}